\documentclass[apj]{emulateapj}
\usepackage{natbib}
\bibstyle{natbib}
\usepackage{graphicx}
\usepackage{amssymb}
\usepackage{amsmath}
\usepackage{mathrsfs}
\usepackage{blindtext}
\usepackage{subfigure}

\newcommand{\beq}{\begin{equation}}
\newcommand{\eeq}{\end{equation}}

\newcommand{\beqary}{\begin{eqnarray*}}
\newcommand{\eeqary}{\end{eqnarray*}}
\newcommand{\h}{\hspace{3mm}}

\begin{document}
\title{Nuclear Mixing Meters for Classical Novae\\}
\author{Keegan J. Kelly\altaffilmark{1,2}, Christian Iliadis\altaffilmark{1,2}, Lori Downen\altaffilmark{1,2}, Jordi Jos\'{e}\altaffilmark{3}, and Art Champagne$^{1,2}$}
\altaffiltext{1}{Department of Physics and Astronomy, University of North Carolina at Chapel Hill, Chapel Hill, NC 27599-3255, USA}
\altaffiltext{2}{Triangle Universities Nuclear Laboratory, Durham, NC 27708-0308, USA}
\altaffiltext{3}{Departament de F{\'i}sica i Enginyeria Nuclear, EUETIB, Universitat Polit\`ecnica de Catalunya, E-08036 Barcelona, Spain and Institut d'Estudis Espacials de Catalunya}

\begin{abstract}
Classical novae are caused by mass transfer episodes from a main sequence star onto a white dwarf via Roche lobe overflow.  This material possesses angular momentum and forms an accretion disk around the white dwarf. Ultimately, a fraction of this material spirals in and piles up on the white dwarf surface under electron-degenerate conditions.  The subsequently occurring thermonuclear runaway reaches hundreds of megakelvin and explosively ejects matter into the interstellar medium.  The exact peak temperature strongly depends on the underlying white dwarf mass, the accreted mass and metallicity, and the initial white dwarf luminosity.  Observations of elemental abundance enrichments in these classical nova events imply that the ejected matter consists not only of processed solar material from the main sequence partner but also of material from the outer layers of the underlying white dwarf.  This indicates that white dwarf and accreted matter mix prior to the thermonuclear runaway.  The processes by which this mixing occurs require further investigation to be understood.  In this work, we analyze elemental abundances ejected from hydrodynamic nova models in search of elemental abundance ratios that are useful indicators of the total amount of mixing.  We identify the abundance ratios $\Sigma$CNO/H, Ne/H, Mg/H, Al/H, and Si/H as useful mixing meters in ONe novae.  The impact of thermonuclear reaction rate uncertainties on the mixing meters is investigated using Monte Carlo post-processing network calculations with temperature-density evolutions of all mass zones computed by the hydrodynamic models. We find that the current uncertainties in the $^{30}$P($p$,$\gamma$)$^{31}$S rate influence the Si/H abundance ratio, but overall the mixing meters found here are robust against nuclear physics uncertainties. A comparison of our results with observations of ONe novae provides strong constraints for classical nova models.
\end{abstract} 


\section{INTRODUCTION}
\label{Intro}
The commonly accepted theory of classical novae involves a white dwarf of carbon-oxygen (CO) or oxygen-neon (ONe) composition accreting matter from a main sequence partner via Roche lobe overflow. The transferred matter carries angular momentum and thus forms an accretion disk. Subsequently, matter accumulates on the surface of the white dwarf under degenerate conditions \citep{Starrfield_1972,Prialnik_1978}.  Once explosive conditions are met a thermonuclear runaway occurs, leading to a violent expulsion of matter \citep{Jose_2006,Starrfield_CN}.\\
\indent Several observables in nova systems (e.g., the light curves and the chemical composition of the ejecta) can only be explained by assuming that matter from the outer layers of the underlying white dwarf is mixed with the accreted matter during or prior to the thermonuclear runaway. This mixing ensures that a sufficient number of carbon, nitrogen, and oxygen nuclei can act as catalysts in the explosive hydrogen burning via the CNO cycles.  Furthermore, spectroscopic observations of large neon abundances for the most energetic classical novae, neon novae, point directly to mixing between accreted matter and a white dwarf of ONe composition.  Numerous physical causes of this mixing have been explored, including diffusion-induced mixing \citep{Prialnik_1984,Kovetz_1985,Iben_1991,Iben_1992,Fujimoto_1992}, shear mixing at the disk-envelope interface \citep{Durisen_1977,Kippenhahn_1978,MacDonald_1983,Livio_1987,Kutter_1987,Sparks_1987}, convective mixing at the core-envelope interface \citep{Woosley_1986,Glasner_1995,Glasner_1997,Glasner_2005,Glasner_2007,Kercek_1998,Kercek_1999,Casanova_2010,Casanova_2011,3DKH}, and mixing by gravity wave breaking on the white dwarf surface \citep{Rosner_2001,Alexakis_2004}.  However, the processes by which outer white dwarf material is mixed with the accreted envelope require further investigation.  \\
\indent Using one-dimensional (1D) hydrodynamic models, \citet{Lori_Nova} recently investigated if ratios of observed elemental abundances can be used to constrain the peak temperature during the explosion. It was found that a number of elemental ratios, including N/O, N/Al, O/S, and S/Al, show a strong monotonic dependence on the peak temperature, and thus represent useful thermometers for the explosion. Since mixing processes cannot be studied self-consistently in a 1D stellar model, a value of 50\% for the degree of mixing between the white dwarf and the accreted envelope was artificially introduced prior to the explosion. This particular value is commonly used in the literature \citep{Politano_1995,Jose_1998,Smith_2002}, but so far, has not been systematically constrained from spectroscopic observations. \\
\indent  It could be argued that the observed overall metallicity, $Z$, of classical novae provides exactly such a constraint. However, there are several reasons to doubt the reliability of reported metallicity values. For example, the review by \citet{Gehrz_1998} lists the following metallicity values obtained by different groups for the same classical nova: $Z=0.39-0.66$ for V693 CrA; $Z=0.10-0.44$ for QU Vul; and $Z=0.18-0.49$ for V1974 Cyg. Therefore, the reported overall metallicity is not a reliable basis for estimating the degree of mixing between the outer white dwarf core and the accreted envelope. We will discuss this issue in more detail below.  \\
\indent Rather than considering the overall metallicity, in the present work we investigate if observed elemental abundance {\it ratios} in classical novae can be used to constrain the degree of mixing between the outer white dwarf layers and the accreted envelope. The results will hopefully shed light on the degree of mixing that occurs and perhaps the physical mixing mechanism responsible, although no definitive conclusions regarding the latter are given in this work. Clearly, any useful mixing meters should exhibit a steep monotonic dependence  on the mixing fraction and, at the same time, they should be insensitive to the white dwarf mass (and hence peak temperature) during the explosion. \\
\indent A number of observations hint at useful mixing meters. First, numerous investigations have shown \citep[see, e.g.,][]{Jose_1998,Yaron_2005,Starrfield_2009,Denissenkov_2013} that during explosive hydrogen burning in classical novae the peak temperatures are less than 400 MK. In this regime, reactions that can bridge the A$<$20 (CNO) and A$\leq$20 mass regions are very slow. Therefore, the total number of CNO nuclei will stay approximately constant during the explosion, and thus the CNO elemental abundance should represent a possible mixing meter \citep[see, e.g.,][]{Kovetz_1997,Jose_1998,Starrfield_2009}. Second, the thermonuclear rate of the $^{20}$Ne($p$,$\gamma$)$^{21}$Na reaction is very slow at classical nova temperatures. Thus, most of the initial $^{20}$Ne nuclei that are mixed  from the white dwarf into the accreted envelope survive the thermonuclear runaway. In fact, this survival of $^{20}$Ne is what first enabled the discovery of ONe novae via emission of the Ne II line \citep{Williams_1985}. Since $^{20}$Ne is expected to be the dominant neon isotope, the observed elemental neon abundance should be a promising mixing meter. \\
\indent We present a systematic investigation of all elements, including CNO and neon, with abundances that can be used to constrain the degree of mixing between matter from the outer white dwarf layers and the accreted envelope in classical novae.  We will focus on ONe novae because these objects display a greater variety of nuclear activity than CO novae.  Also, more reliable spectroscopic abundance data exist for the former class \citep[for the latest elemental abundance compilation, see][]{Lori_Nova}. Results  obtained using a 1D hydrodynamic model are presented in Section \ref{HMC}. Post-processing network calculations are employed to assess the impact of current reaction rate uncertainties and are discussed in Section \ref{PPC}.  Comparisons to observed novae are discussed in Section \ref{CtoO}, and concluding remarks are given in Section \ref{Conclusions}.

\section{HYDRODYNAMIC SIMULATIONS}
\label{HMC}

\begin{figure*}
	\centering
	\includegraphics[width=0.75\textwidth]{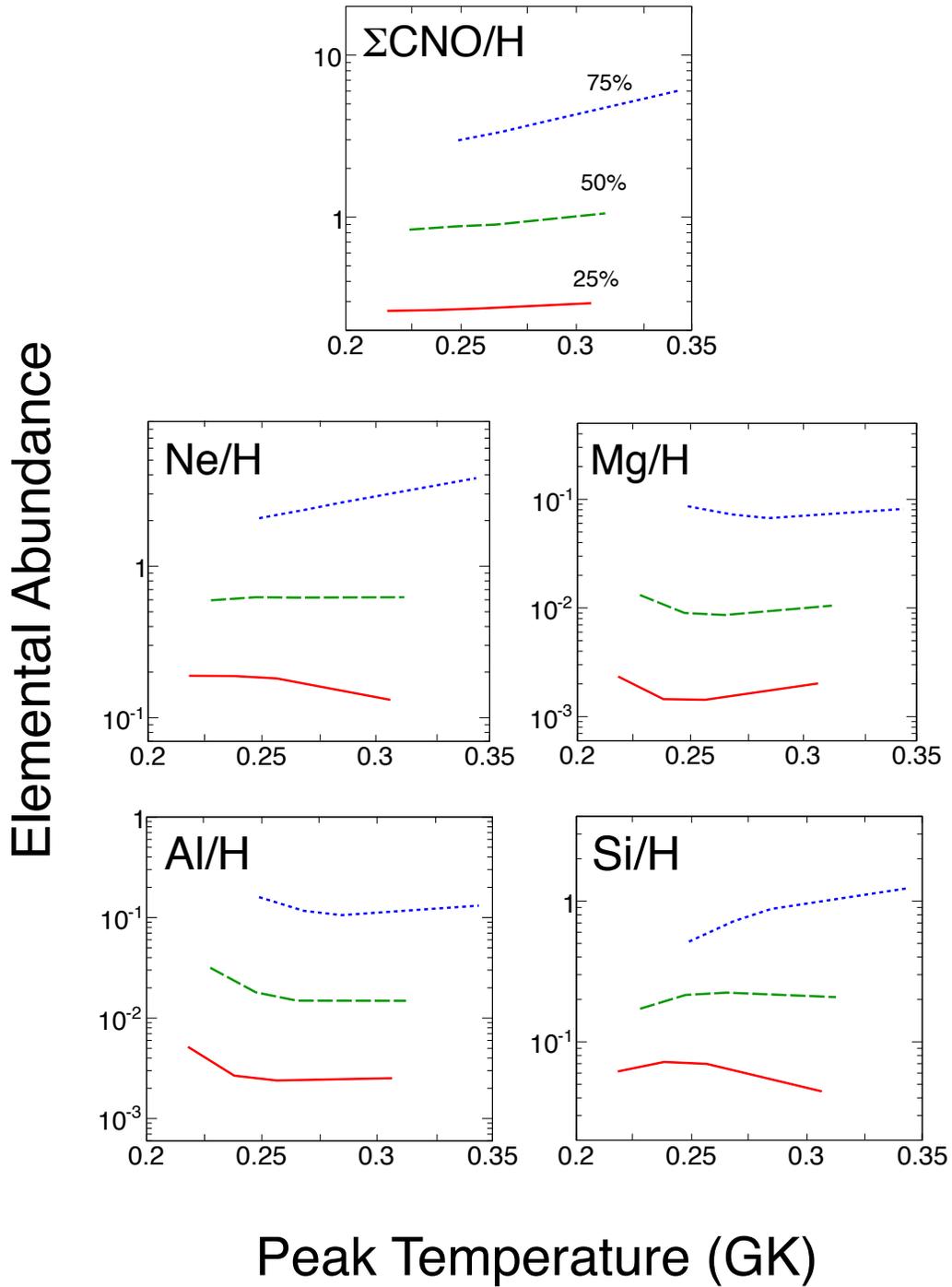}
	\caption{\small Peak temperature versus final elemental abundance (mass fraction) relative to hydrogen for the most promising indicators of mixing between the outer layers of the white dwarf and the accreted envelope.  Results are obtained using the 1D hydrodynamic code SHIVA \citep{Jose_1998}.  The different colors and line styles in each panel correspond to different mixing fractions (indicated by the mass fraction weight of white dwarf material; Blue dotted lines, green dashed lines, and red solid lines indicate 75\%, 50\%, and 25\% white dwarf material by mass, respectively).  Notice the strong dependence of the elemental abundance ratios on the mixing fraction and the small variations with respect to peak temperature.}
	\label{MeterPlotFigure}
\end{figure*}

\begin{figure}
\centering
\includegraphics[width = 0.45\textwidth]{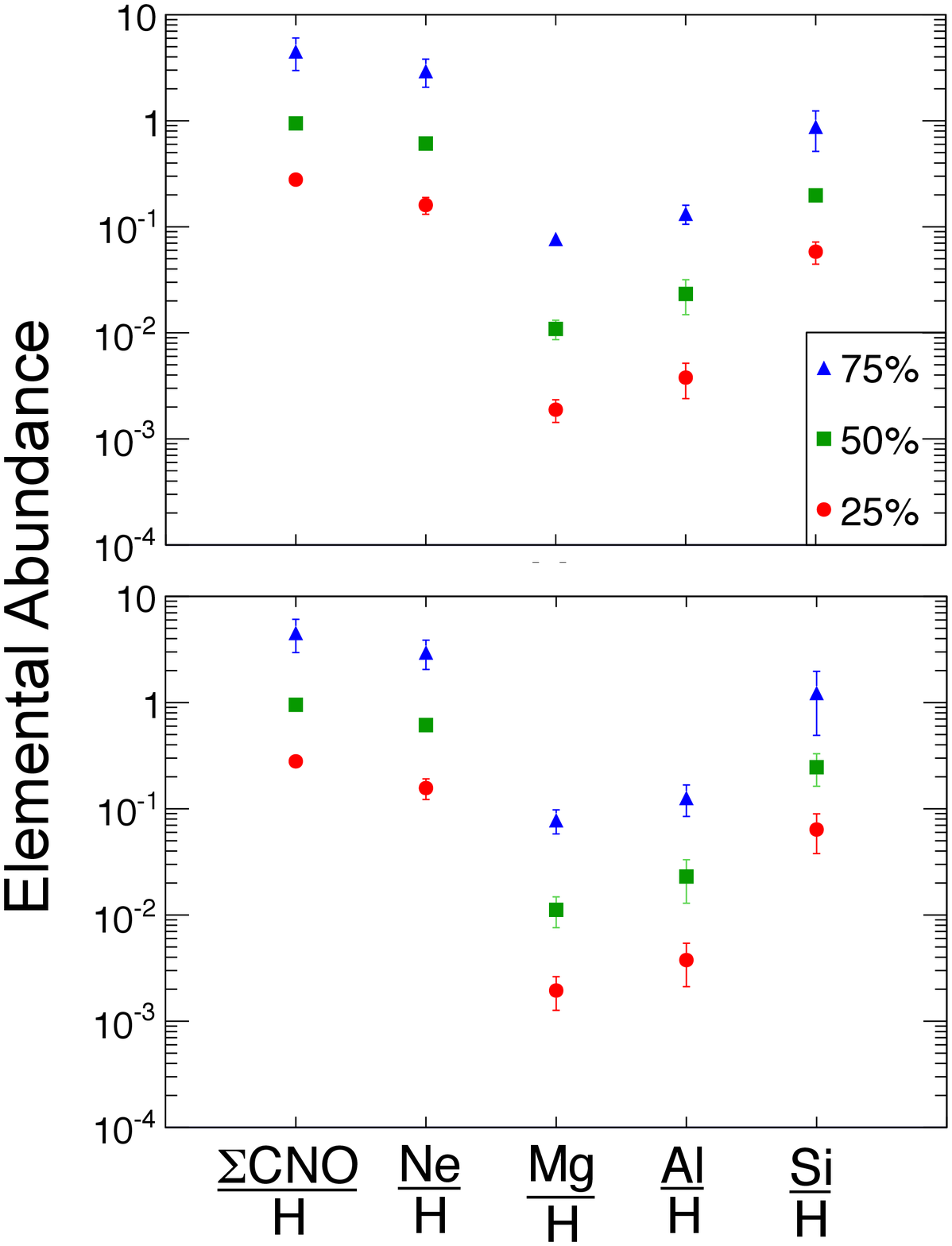}
\caption{\small Elemental abundance (mass fraction) relative to hydrogen for the most promising mixing meters. Different colors and symbols correspond to models computed with different mixing fractions.  (see caption of Figure \ref{MeterPlotFigure}).  Blue triangles, green squares, and red circles indicate mixing fractions of 75\%, 50\%, and 25\%, respectively.  (Top)  Error bars only take the abundance variation with peak temperature into account. (Bottom) Error bars represent abundance variation caused both by differences in white dwarf mass (i.e., peak temperature) and by thermonuclear reaction rate changes.}
\label{Summary}
\end{figure}

\indent Eight new hydrodynamic models were generated using the 1D code SHIVA \citep{Jose_1998} and added to four previously existing models computed with the same code \citep{Lori_Nova}. The models encompass four different underlying white dwarf masses (1.15 M$_\odot$, 1.25 M$_\odot$, 1.30 M$_\odot$, and 1.35 M$_\odot$) and three different mixing fractions between the outer white dwarf layers and the accreted envelope prior to the thermonuclear runaway (25\%, 50\%, and 75\%; the mixing fraction is defined as the weight by mass of the outer white dwarf matter that has been mixed into the envelope prior to nuclear burning). The initial luminosity and mass accretion rate for all models amount to $L_{ini} = 10^{-2}$ L$_\odot$ and $\dot{M}_{acc} = 2 \times 10^{-10}$ M$_\odot$ yr$^{-1}$, respectively. Information on the model parameters (peak temperature, mixing fraction, and ejected mass) is summarized in Table \ref{novaChars}. \\
\indent The nuclear reaction network consists of 117 nuclides ranging from $^1$H to $^{48}$Ti, linked by 654 interactions, including ($p$,$\gamma$), ($p$,$\alpha$), and ($\alpha$,$\gamma$) reactions and weak decays. The nuclear interaction rates were adopted from recommended values provided by the STARLIB library \citep{Anne_Starlib}. For most reactions of interest to classical novae, the experimental rates contained in STARLIB have been obtained by the Monte Carlo method described in \citet{Longland_I_2010} and \citet{Iliadis_II_2010}. The library also provides the rate probability density function at temperatures in a grid ranging from 10 MK to 10 GK. This unique feature will be important for the post-processing studies, as discussed in more detail in Section \ref{PPC}.\\
\indent We assume that matter is accreted from a main sequence companion of solar composition \citep{Lodders_2009}. For the composition of the underlying white dwarf we adopt the results obtained from the evolution of a 10 M$_\odot$ star from the main sequence to the end of core carbon burning \citep{Ritossa_1996}. The initial envelope composition for different mixing fractions  of our nova models is presented in Table \ref{initAbs}.  Each model includes 45 envelope zones encompassing all material involved in the thermonuclear runaway. Test calculations performed with 500 zones provide essentially the same results. The peak temperatures achieved in the hottest zone in our simulations, ranging from 218 MK to 344 MK depending on the model, can be regarded as typical for models of thermonuclear runaways involving ONe white dwarfs. \\
\indent Final isotopic abundances for matter that exceeds escape velocity (i.e., the fraction of the envelope effectively ejected) are determined 1 hour after peak temperature is achieved. By that time, short-lived parent nuclei would have decayed to their stable daughters.  Isotopic abundances of a given element are summed for each model. Elemental abundances that reveal a strong, monotonic dependence on the mixing between the outer white dwarf layers and the accreted envelope, and at the same time are insensitive to the white dwarf mass (or peak temperature), are selected for further inspection. Following the procedure applied in \citet{Lori_Nova}, we focus on elemental mass fractions relative to hydrogen, $X_{el}/X_H$, instead of elemental mass fractions, $X_{el}$, since observed classical nova abundances are usually determined relative to hydrogen (see Section \ref{CtoO}).\\
\indent Figure \ref{MeterPlotFigure} shows the computed final elemental abundance relative to hydrogen versus the peak temperature for the most promising mixing meters. The different colors and line styles in each panel correspond to different mixing fractions (see Table \ref{initAbs}). It is apparent that all displayed abundance ratios depend only weakly on peak temperature (by about a factor of 2 at most), while revealing a strong, monotonic dependence on the mixing fraction (by at least an order of magnitude). As expected, the values of both $\Sigma$CNO/H and Ne/H are useful mixing meters. In addition, we find that the ratios Mg/H, Al/H, and Si/H satisfy the necessary conditions for a mixing meter.  The ratio Na/H also satisfies these criteria but was disregarded here because no observational Na data exist yet for ONe novae.\\
\indent The results are summarized in the top panel of Figure \ref{Summary}, showing the computed final elemental abundance ratios of these five mixing meters. Different symbols and colors correspond to different mixing fractions (see Table \ref{initAbs}). The error bar on each data point represents the abundance change when the peak temperature is varied across our models for a fixed mixing fraction (see Table \ref{novaChars}). The important point to emphasize is that, for a given elemental ratio, the abundance values vary by 1-2 orders of magnitude with mixing fraction, and the error bars (caused by peak temperature changes alone) do not overlap.  The impact of thermonuclear reaction rate variations on the mixing meter values will be addressed in the following section.\\

\begin{deluxetable}{cccc}
\centering
\tabletypesize{\footnotesize}
\tablewidth{0pt}
\tablecolumns{4}
\tablecaption{Selected Results for Characteristics of Hydrodynamic ONe Nova Models}
\tablehead{
\colhead{Mixing Fraction}& \colhead{$\h $M$_{WD}($M$_{\odot})\h$} & \colhead{T$_{peak}$(GK)$\h$} & \colhead{M$_{ej}(10^{-5}$M$_{\odot})$}}
\startdata
\hline\\
25\%&$1.15 $&$0.218$&2.12	\\
&$1.25 $&$0.238$&1.70	\\
&$1.30 $&$0.256$&1.18	\\
&$1.35 $&$0.306$&0.429\\	
\hline\\
50\%&$1.15$&$0.228$&2.46	\\
&$1.25$&$0.247$&1.89	\\
&$1.30$&$0.265$&1.17	\\
&$1.35$&$0.316$&0.455	\\
\hline\\
75\%&$1.15$&$0.249$&2.44	\\
&$1.25$&$0.268$&1.88	\\
&$1.30$&$0.284$&1.29	\\
&$1.35$&$0.344$&0.447	
\enddata
\label{novaChars}
\tablecomments{The models were computed using the 1D hydrodynamic code SHIVA \citep{Jose_1998}.  The mixing fraction is defined as the  weight by mass of the outer white dwarf matter that has been mixed into the envelope prior to nuclear burning.}
\end{deluxetable}

\begin{deluxetable}{cccc}
\tabletypesize{\footnotesize}
\tablewidth{0pt}
\tablecaption{Initial Abundances for Hydrodynamic Calculations}
\tablehead{
\colhead{Nuclide} & & Mass Fraction &\\
\cline{2-4}
&  \colhead{$25\%^{\dagger}$} & \colhead{$50\%^{\dagger}$} & \colhead{$75\%^{\dagger}$}
}
\startdata
$	^	{	1	}$H		&	5.33E-01	&	3.53E-01	&	1.78E-01	\\
$	^	{	3	}$He		&	6.35E-05	&	2.62E-05	&	2.12E-05	\\
$	^	{	4	}$He		&	2.05E-01	&	1.38E-01	&	6.84E-02	\\
$	^	{	6	}$Li		&	5.16E-10	&	3.25E-10	&	1.72E-10	\\
$	^	{	7	}$Li		&	7.37E-09	&	4.68E-09	&	2.46E-09	\\
$	^	{	9	}$Be		&	1.13E-10	&	8.31E-11	&	3.76E-11	\\
$	^	{	10	}$B		&	7.57E-10	&	5.33E-10	&	2.52E-10	\\
$	^	{	11	}$B		&	3.37E-09	&	2.36E-09	&	1.13E-09	\\
$	^	{	12	}$C		&	4.03E-03	&	6.10E-03	&	7.45E-03	\\
$	^	{	13	}$C		&	2.12E-05	&	1.82E-05	&	7.07E-06	\\
$	^	{	14	}$N		&	6.07E-04	&	5.53E-04	&	2.02E-04	\\
$	^	{	15	}$N		&	2.38E-06	&	2.18E-06	&	7.95E-07	\\
$	^	{	16	}$O		&	1.33E-01	&	2.61E-01	&	3.85E-01	\\
$	^	{	17	}$O		&	2.05E-06	&	1.94E-06	&	6.83E-07	\\
$	^	{	18	}$O		&	1.16E-05	&	1.08E-05	&	3.86E-06	\\
$	^	{	19	}$F		&	3.12E-07	&	2.03E-07	&	1.04E-07	\\
$	^	{	20	}$Ne		&	7.95E-02	&	1.57E-01	&	2.35E-01	\\
$	^	{	21	}$Ne		&	1.50E-03	&	2.99E-03	&	4.49E-03	\\
$	^	{	22	}$Ne		&	1.18E-03	&	2.22E-03	&	3.27E-03	\\
$	^	{	23	}$Na		&	1.61E-02	&	3.22E-02	&	4.83E-02	\\
$	^	{	24	}$Mg		&	1.41E-02	&	2.77E-02	&	4.12E-02	\\
$	^	{	25	}$Mg		&	4.00E-03	&	7.93E-03	&	1.19E-02	\\
$	^	{	26	}$Mg		&	2.53E-03	&	4.98E-03	&	7.44E-03	\\
$	^	{	27	}$Al		&	2.75E-03	&	5.43E-03	&	8.12E-03	\\
$	^	{	28	}$Si		&	5.27E-04	&	3.27E-04	&	1.76E-04	\\
$	^	{	29	}$Si		&	1.83E-05	&	1.71E-05	&	6.10E-06	\\
$	^	{	30	}$Si		&	1.89E-05	&	1.18E-05	&	6.30E-06	\\
$	^	{	31	}$P		&	5.25E-06	&	4.08E-06	&	1.75E-06	\\
$	^	{	32	}$S		&	2.61E-04	&	1.98E-04	&	8.71E-05	\\
$	^	{	33	}$S		&	2.13E-06	&	1.61E-06	&	7.09E-07	\\
$	^	{	34	}$Cl		&	2.80E-06	&	1.27E-06	&	9.33E-07	\\
$	^	{	37	}$Cl		&	9.49E-07	&	4.27E-07	&	3.15E-07	\\
$	^	{	36	}$Ar		&	5.76E-05	&	3.87E-05	&	1.92E-05	\\
$	^	{	38	}$Ar		&	1.11E-05	&	7.69E-06	&	3.70E-06	\\
$	^	{	39	}$K		&	2.79E-06	&	1.73E-06	&	9.29E-07	\\
$	^	{	40	}$Ca		&	4.78E-05	&	2.99E-05	&	1.59E-05	\\
\enddata
\tablecomments{For the outer layers of the ONe white dwarf, the abundances are taken from \citet{Ritossa_1996}; solar abundances for the accreted matter are adopted from \citet{Lodders_2009}.\\
$^{\dagger}$Mixing fraction, defined as the weight by mass of the outer white dwarf matter that has been mixed into the envelope prior to nuclear burning.}
\label{initAbs}
\end{deluxetable}


\section{POST-PROCESSING CALCULATIONS}
\label{PPC}
\begin{figure}
	\centering
	\includegraphics[width=0.45\textwidth]{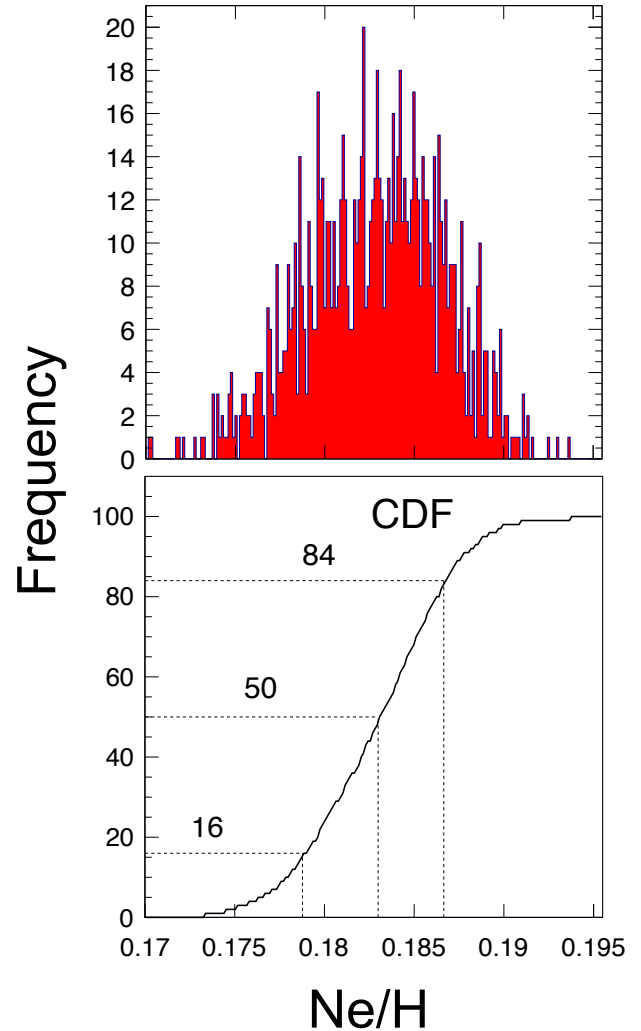}
	\caption{\small{(Top) Final elemental abundance (mass fraction) ratio probability density for Ne/H, obtained from our Monte Carlo post-processing nucleosynthesis study by simultaneously sampling all reaction rates in our network (1000 samples).  The calculations are performed with the temperature-density history derived from a hydrodynamic model (1.30 M$_\odot$ ONe white dwarf, 25\% mixing fraction).  The post-processing calculations take into account all envelope zones.  (Bottom) Corresponding cumulative distribution function (CDF).  The 16th, 50th, and 84th percentiles of the final elemental Ne/H abundance amount to 0.179, 0.183, and 0.187, respectively, resulting in X$_{Ne}$/X$_H$=0.183$\pm$0.004 (for a coverage probability of 68\%).}}
	\label{NucUnc}
	\end{figure}

\begin{figure*}
	\centering
	\includegraphics[width=.45\textwidth,angle=90]{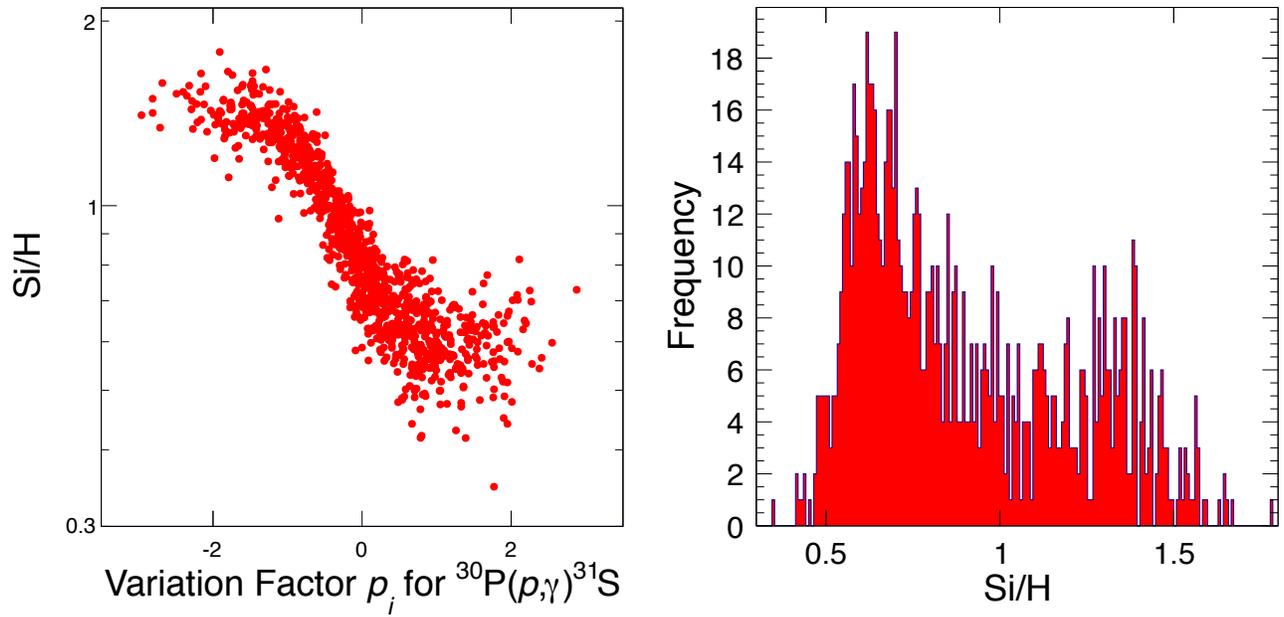}
	\caption{\small (Left)  Correlation of final elemental Si/H mass fraction ratio with the $^{30}$P($p$,$\gamma$)$^{31}$S reaction rate for a nova model with a 1.35 M$_\odot$ ONe white dwarf with 75\% mixing.  (Right)  Final elemental Si/H mass fraction ratio distribution for the same model.  This reaction rate is a leading cause of the uncertainty shown in the bottom panel of Fig \ref{Summary}.}
	\label{SiH_30Ppg}
\end{figure*}

\indent We will now discuss how reaction rate uncertainties influence the abundance ratios of $\Sigma$CNO/H, Ne/H, Mg/H, Al/H, and Si/H. If variation of reaction rates within uncertainties gives rise to large changes in these abundance ratios, then they are clearly not useful as  mixing meters. Since a large number of calculations need to be executed, these sensitivity studies are performed by post-processing the temperature-density versus time profiles extracted from the hydrodynamic simulations. Similar to the procedure of \citet{Lori_Nova}, we do not only take into account the hottest zone, but all envelope zones. Furthermore, we adopt the simplified assumptions for convection of \citet{Lori_Nova}, assuming either instantaneous mixing, no mixing, or the geometric mean of instant and no mixing between zones based on which description best reproduces the final abundances of the hydrodynamic model. This procedure works well and reproduces the final abundances of the hydrodynamic models within a factor of 2.  But unlike their procedure of varying one nuclear reaction rate at a time within uncertainties\footnote{In \citet{Lori_Nova}, the rate variation for each nuclear reaction was explored by performing three post-processing calculations, by using the low rate, recommended rate, and high rate. These rates are defined by the 16th, 50th, and 84th percentile, respectively, of the cumulative rate distribution. The Monte Carlo method applied in the present work not only samples over the complete rate probability density (instead of 3 values only), but it also takes the abundance changes caused by the simultaneous variation of all rates into account.}, we employ here a Monte Carlo technique. This method takes advantage of unique properties of the STARLIB reaction rate library \citep{Anne_Starlib}, as described below.\\
\indent The majority of experimental thermonuclear reaction rates important for classical nova nucleosynthesis included in STARLIB \citep{Anne_Starlib} were obtained using a Monte Carlo method \citep{Longland_I_2010,Iliadis_II_2010}. In brief, reaction rates are randomly sampled many times over the input nuclear physics probability density functions. The (output) reaction rate probability density is then used to define a statistically meaningful recommended rate as the 50th percentile of the cumulative rate distribution. These are the values that were employed for the hydrodynamic calculations discussed in the previous section. \\
\indent \citet{Longland_I_2010} found that most reaction rate probability densities follow a lognormal distribution. This distribution has two parameters, the location parameter $\mu$ and, the spread parameter $\sigma$, which are also tabulated in STARLIB. With the tabulated values, the rate probability density can be calculated at all temperatures of interest. This feature is very useful for a Monte Carlo nucleosynthesis simulation, where all rates are simultaneously sampled  according to their individual probability densities \citep{Longland_2012}.  Rate samples are drawn according to the function \citep{Anne_Starlib}: 
\beqary
x_i = e^{\mu ^{_{+}} p_i\sigma} = e^{\mu}(e^{\sigma})^{p_i} = x_{med}(f.u.)^{p_i}
\eeqary
where $x_i$ denotes the sampled rate, $e^{\mu} = x_{med}$ is the median rate, $e^\sigma$ is the rate uncertainty factor, and $p_i$ is a normally distributed random number with a mean of zero and standard deviation of unity. For a random sample of $p_i=0$, the recommended rate ($e^\mu$) is obtained. All the rates are simultaneously sampled once at the beginning of each multi-zone post-processing calculation, i.e., the probability factor $p_i$ is sampled independently for each reaction, $i$, in the network. For a given reaction, the quantity $p_i$ has the same value at all temperatures. This method corresponds to the flat parametrization introduced by \citet{Longland_2012}. Notice that, for a fixed value of $p_i$, the reaction rate is temperature-dependent through the (tabulated) lognormal parameter $\sigma$. The Monte Carlo procedure generates distributions of final elemental abundances, and the spread of these distributions indicates the abundance uncertainty caused by simultaneously varying all reaction rates. Furthermore, the parameters $p_i$ are saved for each reaction network run in order to study correlations between specific reaction rates and between reaction rates and final abundances.\\
\indent In total, 1000 nuclear reaction network samples were computed for each of the twelve combinations of the white dwarf mass and mixing fraction (see Table \ref{novaChars}).  This number of samples was found to reproduce median abundances to within 5\% \citep[see also][]{Longland_2012}.  As already noted above, the median final elemental abundance values from our Monte Carlo post-processing calculations, derived from the 50th percentile of the cumulative abundance distribution, agree with the final abundances from the \emph{hydrodynamic} models within a factor of 2. Therefore, the abundance uncertainties derived from the post-processing Monte Carlo procedure satisfactorily reflect the impact of current thermonuclear reaction rate uncertainties. Final abundance uncertainties are derived from the 16th and 84th percentiles of the cumulative distribution functions (for a coverage probability of 68\%).  As an example, we show the Monte Carlo results for the final Ne/H abundance ratio in the top panel of Figure \ref{NucUnc}, obtained for a 1.30 M$_\odot$ ONe white dwarf and a mixing fraction of 25\%. The corresponding cumulative abundance distribution is shown in the bottom panel. The final Ne/H mass fraction ratio amounts to 0.183$\pm$0.004, in good agreement with the Ne/H value obtained from the hydrodynamic model (0.182).  \\
\indent Results for all five mixing meters, $\Sigma$CNO/H, Ne/H, Mg/H, Al/H, and Si/H, are displayed in the bottom panel of Figure \ref{Summary}, where the error bars represent the abundance variation caused both by differences in peak temperature {\it and} thermonuclear reaction rate changes. Comparison to the top panel, which only reflects the variation of each mixing meter abundance based on white dwarf mass (i.e., peak temperature), reveals that the uncertainties caused by thermonuclear reaction rates are relatively small. \\
\indent The Si/H abundance ratio shows the largest impact of reaction rate variations, mainly caused by a dependence on the $^{30}$P($p$,$\gamma$)$^{31}$S reaction rate.  This reaction involves a short-lived nuclide (t$_{1/2}$ = 2.498 minutes) and has not been measured directly yet.  Its importance for classical nova nucleosynthesis was pointed out by \citet{Jose_2001}.  Indirect nuclear structure studies have been reported \citep{Wrede_2009,Parikh_2011,Doherty_2012}. At present, the spin and parity assignments for some of the $^{31}$S proton threshold states are ambiguous. In addition, none of the proton partial widths for these levels are experimentally known. For this reason, the rate of this reaction listed in STARLIB was calculated using the Hauser-Feshbach model \citep{Anne_Starlib}, assuming a factor of 10 uncertainty within the range of temperatures characteristic for classical novae. Our adopted rates for this reaction agree within their uncertainties with those of \citet{Parikh_2011} and \citet{Doherty_2012}, both of which were obtained using different procedures.  The correlation of the final Si/H elemental abundance ratio with the rate of the $^{30}$P($p$,$\gamma$)$^{31}$S reaction along with the corresponding elemental abundance distribution is shown in the left panel of Figure \ref{SiH_30Ppg} for a 1.35 M$_\odot$ white dwarf and a 75\% mixing fraction.  It can be seen that a $^{30}$P($p$,$\gamma$)$^{31}$S reaction rate increase causes the Si/H ratio to decrease.  This is to be expected because as $^{30}$P is destroyed via proton capture, silicon production via $^{30}$P($\beta^+ \nu$)$^{30}$Si becomes less likely.  \\
\indent It must be emphasized that, for a given abundance ratio, none of the error bars in the bottom panel of Figure 2 overlap. This point is important since it demonstrates that the five mixing meters are robust with regard to thermonuclear reaction rate variations.

\clearpage
\section{COMPARISON TO OBSERVATIONS}
\label{CtoO}

\begin{figure*}
	\centering
	\includegraphics[width=.85\textwidth]{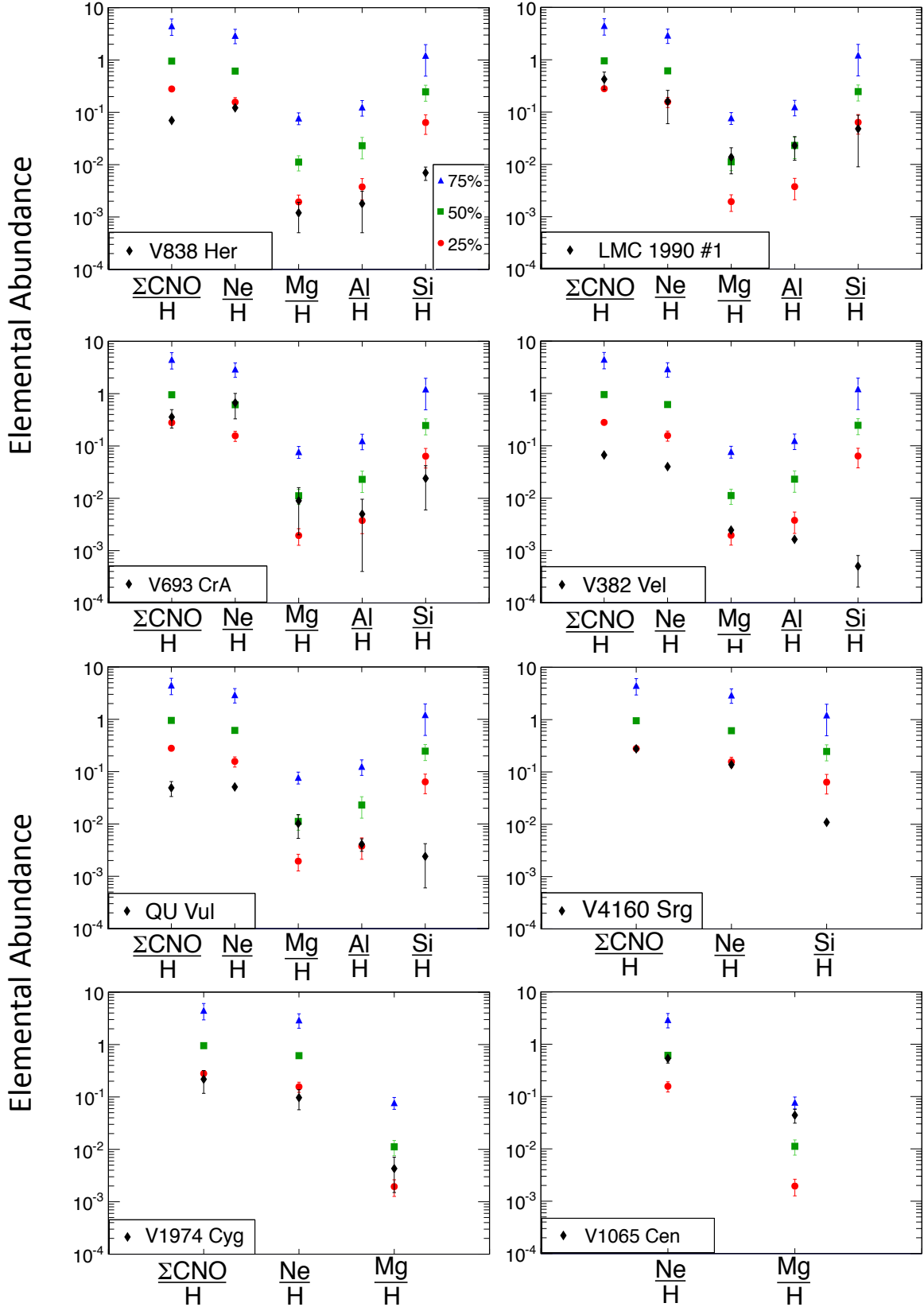}
	\caption{\small (Color online) Comparison of observed ONe nova elemental abundances (black data points) with our set of suggested mixing meters.  The colored points indicate our model calculations and are the same as those shown in the bottom panel of Figure \ref{Summary}. Different colors and symbols denote different mixing fractions: 75\% (blue triangles); 50\% (green squares); 25\% (red circles). Error bars on the calculated points include the impact of variations in both peak temperature and thermonuclear reaction rates.}
\label{NovaComp}
\end{figure*}

\indent We will now compare our numerical results to observations. The difficulties in determining classical nova abundances have been discussed in detail before \citep{Jose_CN}.  First, the derived abundances are model dependent since most analyses assume spherical shells although there is evidence that the ejecta are not spherically symmetric \citep{3DKH}.  Second, the ejecta are chemically inhomogeneous, adding substantially to the complexity of the problem.  Third, the filling fraction (i.e., the fraction of the shell volume occupied by gas) is poorly constrained.  Fourth, the abundance analysis must account for the (sometimes substantial) fraction of unobserved ionization states. What is usually directly extracted by fitting nova spectral line intensities is the quantity:
\beq
\xi = \frac{(N_{el}/N_{H})}{(N_{el}/N_{H})}_{\odot}
\eeq
where $N_{el}$ refers to the number abundance. The conversion to a ratio of mass fractions, $X_{el}/X_H$ is straightforward and the most recent compilation of such values is presented in \citet{Lori_Nova}.  As a word of caution, we emphasize that the hydrogen mass fraction can be calculated from
\beq
X_{H} = \left[1+\frac{X_{He}}{X_{H}} + \frac{X_{C}}{X_{H}} + ... + \frac{X_{Fe}}{X_{H}}\right]^{-1}
\eeq
which can be used to estimate absolute mass fractions, $X_{el}$, or metallicities, $Z=1-X_H-X_{He}$. However, the values of $X_{el}$ derived in this manner are sensitive to the abundance fraction missed in the spectral line analysis, such as missing elements or ionization states not accounted for. This was the main reason why we focussed in the present work on the quantity $X_{el}$/$X_H$, which is much less susceptible to systematic errors compared to either $X_{el}$ or $Z$.  \\
\indent Our post-processing results are summarized in Figure \ref{NovaComp} for each observed classical nova. The black markers correspond to the observations, while the colored markers indicate our model calculations (see the bottom panel of Figure \ref{Summary}). Although we explored only three different mixing fractions, a number of interesting observations can be made for specific novae. \\
\indent For nova V838 Her, Ne/H, Mg/H, and Al/H consistently suggest a mixing fraction of 25\%, while Si/H and $\Sigma$CNO/H indicate an even smaller value. In the case of nova LMC 1990 \#1, Ne/H, Si/H, and $\Sigma$CNO/H imply a 25\% mixing fraction, while Mg/H and Al/H suggest an even smaller value. For nova V693 CrA, Mg/H, Al/H, Si/H, and $\Sigma$CNO/H are consistent with a mixing fraction of 25\%; although Ne/H may suggest a value of 50\%, the observational error bar is relatively large.  Novae V382 Vel and V1974 Cyg, which have similar observed abundances, show poor agreement with the mixing meters.  A mixing fraction of less than 25\% is predicted for these novae.  Only three mixing meters are available for novae V4160 Srg and V1974 Cyg; the observations are in good agreement with a mixing fraction of 25\%, although the observed Si/H value is smaller than the predicted one. Finally, only two mixing meters can be applied to nova V1065 Cen, and they indicate a mixing fraction of 50\% or larger. \\
\indent We conclude that the mixing meters investigated in the present work clearly indicate a fraction of 25\% or less for the mixing between white dwarf matter and the accreted envelope. The only exception is nova V1065 Cen, although one must remember that only two mixing meters are available and more observations are warranted before firmer conclusions can be drawn. This suggests that the observations do not support a mixing fraction of 50\%, a value that has usually been used in the literature (on the basis of inferred overall metallicities of the ejecta).\\
\indent Unfortunately, there are no predicted mixing fractions for individual ONe novae reported in the literature. However, our results are consistent with some predictions for CO models: \citet{Glasner_2007} reported a range of 35\%$-$50\% mixing from investigations of convective undershoot mixing on a 1.14 M$_{\odot}$ CO white dwarf; \citet{Kovetz_1985} presented $\Sigma$CNO of 0.08$-$0.4 (presumably implying an overall mixing fraction of 8\% to 40\%) based on diffusion layer mixing in novae involving CO white dwarfs ranging from 0.9 M$_{\odot}$ to 1.25 M$_{\odot}$; \citet{Fujimoto_1988} determined a ри modest degree of mixing ис from theoretical calculations of elemental mixing due to differential rotation. Although two of the above sources are specific to CO novae, the results qualitatively agree with the mixing fractions suggested in Figure \ref{NovaComp}. 
\section{CONCLUSIONS}
\label{Conclusions}
\indent We investigated if observed ONe nova abundances can be used to constrain the degree of mixing that occurs between the outer layers of the underlying white dwarf and the accreted envelope prior to thermonuclear runaway.  Any abundance used for this purpose, referred to as a mixing meter, should show a steep monotonic dependence on the mixing fraction, but at the same time, it should be insensitive to peak temperature (i.e. white dwarf mass). By performing hydrodynamic model calculations with the latest thermonuclear reaction rates, we identified the elemental abundance ratios $\Sigma$CNO/H, Ne/H, Mg/H, Al/H, and Si/H  as useful mixing meters. The impact of thermonuclear reaction rate uncertainties on these abundances was explored using Monte Carlo post-processing reaction network calculations.  We demonstrated that reaction rate uncertainties have only a small influence on the mixing meters, meaning the mixing meters are robust with respect to nuclear physics uncertainties.\\
\indent Comparison of mixing meters to observations allowed for an estimate of the mixing fractions in individual novae. We find a fraction of 25\% or smaller for the mixing between white dwarf matter and the accreted envelope in almost all cases (ONe models).  Therefore, the observations  support a mixing fraction that is much smaller compared to the value of 50\%, which has usually been used in the literature. \\
\\
\\
\indent  This work was supported in part by the US Department of Energy under grant DE-FG02-97ER41041, the National Science Foundation under award number AST-1008355, the Spanish MICINN grants AYA2010-15685 and EUI2009-04167, the E. U. FEDER funds, and the ESF EUROCORES Program EuroGENESIS.

\end{document}